\documentstyle[PASJadd,graphicx]{PASJ95}

\newcommand{\vol}{}
\newcommand{\no}{}
\newcommand{\lett}{}
\newcommand{\spage}{}
\newcommand{\rdate}{}
\newcommand{\adate}{}

\begin{document}

\title{Dual-Frequency VSOP Observations of AO 0235+164}

\author{S\'andor {\sc Frey} \\
{\it F\"OMI Satellite Geodetic Observatory, Penc, Hungary} \\
{\it E-mail: frey@sgo.fomi.hu}\\[6pt]
Leonid I. {\sc Gurvits} \\
{\it Joint Institue for VLBI in Europe, Dwingeloo, The Netherlands} \\[6pt]
Daniel R. {\sc Altschuler}, Michael M. {\sc Davis}, Phil {\sc Perillat}, Chris J. {\sc Salter} \\
{\it National Astronomy and Ionosphere Center, Arecibo Observatory, Puerto Rico} \\[6pt]
Hugh D. {\sc Aller}, Margo F. {\sc Aller} \\
{\it Department of Astronomy, University of Michigan, USA} \\
and \\
Hisashi {\sc Hirabayashi} \\
{\it Institute of Space and Astronautical Science, Sagamihara, Japan} \\
}

\abst{AO 0235+164 is a very compact, flat spectrum radio source identified as
a BL Lac object at a redshift of $z=0.94$. It is one of the most violently variable
extragalactic objects at both optical and radio wavelengths. 
The radio structure of the source revealed by various ground-based VLBI 
observations is dominated by a nearly unresolved compact component at 
almost all available frequencies.

Dual-frequency space VLBI observations of AO 0235+164 were made 
with the VSOP mission in January-February 1999.
The array of the Japanese HALCA satellite and co-observing 
ground radio telescopes in Australia, Japan, China and South Africa allowed 
us to study AO 0235+164 with an unprecedented angular resolution at frequencies of 
1.6 and 5~GHz. We report on the sub-milliarcsecond structural properties 
of the source. The 5-GHz observations led to an estimate of 
$T_{\mathrm B}>5.8\times10^{13}$~K for the rest-frame brightness temperature
of the core, which is the highest value measured with VSOP to date. 
}

\kword{Galaxies: active --- Radio continuum: galaxies --- Radio sources: variable 
--- BL Lacertae objects: individual (AO 0235+164) --- Techniques: interferometric}

\maketitle
\thispagestyle{headings}

\section
{Introduction}

The Arecibo Occultation source AO 0235+164 is a well known object extensively studied 
at nearly all wavelengths. It was identified as a BL Lac object by Spinrad and Smith (1975)
based on its almost featureless optical spectrum. Its emission line redshift of $z=0.940$ 
was first measured by Cohen et al. (1987). Two other absorption redshift systems were 
detected in the same direction at $z=0.851$ and $z=0.524$. The latter was detected also 
in emission. There are a number of foreground galaxies within a few arsceconds of AO 0235+164 
at the latter redshift, indicating the presence of a group or cluster of star--forming galaxies. 
Extensive optical studies of the object AO 0235+164 and the surrounding field, both imaging 
and spectroscopy, are presented in Nilsson et al. (1996), Burbidge at al. (1996) 
and references therein. 

The source is highly variable in the radio (O'Dell et al. 1988; Romero et al. 1997), 
near-infrared (Takalo et al. 1992), optical (Rabbette et al. 1996) and X-ray (Madejski et al. 1996)
regimes of the electromagnetic spectrum.  It is detected in gamma-rays by the EGRET on the 
Compton Gamma-Ray Observatory (von Montigny et al. 1995), with indication of variability 
during the 1-year observing period.
The presence of a group of foreground galaxies led Stickel, Fried \& K\"uhr (1988) to suggest 
that the dramatic flux density variations, correlated at optical and radio wavelengths, 
could be explained in terms of gravitational microlensing. Abraham et al. (1993) performed 
deep imaging of AO 0235+164. Based on the observation and modelling its Mg{\sc ii} absorber, 
they concluded that strong microlensing of AO 0235+164 by individual stars in the intervening 
galaxy is unlikely. Saust (1992) came to a similar conclusion using emission line variability 
measurements. It appears that variability is intrinsic to the source and may be best explained 
by the relativistic beaming model.



\begin{table*}[t]
\small
\begin{center}
Table~1.\hspace{4pt}Imaging VLBI observations of AO 0235+164 from the literature.\\
\end{center}
\vspace{6pt}
\begin{tabular*}{\textwidth}{@{\hspace{\tabcolsep}
\extracolsep{\fill}}p{7pc}lllllllll}
\hline\hline\\[-6pt]
$\nu$ & Epoch       & $S_0$ $^a$ & $S^b$  & size$^c$         & $T_{\mathrm B}$ $^d$ & Array$^e$ & Extension PA & Reference$^f$ \\
(GHz) & (year)      & (Jy)    & (Jy)      & (mas$\times$mas) & ($10^{11}$ K) &        &              &           \\
\noalign{\smallskip}
\hline
\noalign{\smallskip}
0.327 & 1986.25     & 0.88    & 0.52      & 19$\times$13     & 0.5           & global & $22^{\circ}$ & AG95 \\
1.6   & 1999.08     & 1.44    & 1.44      & 0.85$\times$0.60 & 18            & SVLBI  & no           & this paper \\
2.3   & 1981.89     & 1.2     & 0.8       &                  &               & global & $35^{\circ}$       & CU83 \\
2.3   & 1996.31     &         & 0.67$^g$  & 7.55$\times$3.75 & 0.1           & VLBA   & E-NE         & USNO \\
2.3   & 1997.03     &         & 0.41$^g$  & 7.03$\times$3.54 & 0.08          & VLBA   & E-NE         & USNO \\
5     & 1978.9      & 1.89    & 1.89      & 3.5$\times$3.5   & 0.1           & Europe & uncertain    & BE81 \\
5     & 1979.2      & 3.36    & 3.11      & 2.55$\times$1.48 & 0.8           & global & $25^{\circ}$ & CB96 \\
5     & 1980.8      & 2.26    & 2.18      & 0.42$\times$0.19 & 26            & global & $14^{\circ}$ & CB96 \\
5     & 1981.9      & 1.78    & 1.72      & 0.31$\times$0.31 & 17            & global & $47^{\circ}$ & CB96 \\
5     & 1981.92     & 2.14    &           &                  &               & US     & no           & GC89 \\
5     & 1982.93     & 1.68    &           &                  &               & US     & no           & GC89 \\
5     & 1983.5      & 2.27    & 1.81      & $<0.1\times<0.1$ & $>172$        & global & $-163^{\circ}$ & CB96 \\
5     & 1986.45     & 1.57    &           &                  &               & global & no           & GC89 \\
5     & 1987.41     & 2.87    & 2.87      & 0.53$\times$0.53 & 10            & global & no           & GC92 \\
5     & 1992.9      & 4.57    & 4.45      & 0.4$\times$0.2   & 53            & global & $58^{\circ}$ & SW97 \\
5     & 1996.42     & 0.43    & 0.43      & $<0.7\times$0.6 & $>10$          & VLBA   & no           & FF00 \\
5     & 1997.13     & 0.30    & 0.29      & 0.79$\times$0.59 & 0.5           & EVN    & $47^{\circ}$ and $-45^{\circ}$ & CZ99 \\
5     & 1999.08     & 1.23    & 0.92      & $<0.05\times<0.03$ & $>580$      & SVLBI  & $-57^{\circ}$ & this paper \\
8.4   & 1990.47     & 2.67    & 2.67      & 0.1$\times$0.1   & 90            & global & no           & GC96 \\
8.4   & 1996.31     &         & 0.29$^g$  & 2.07$\times$1.03 & 0.05          & VLBA   & E and E-NE?  & USNO \\
8.4   & 1997.03     &         & 0.21$^g$  & 1.9$\times$0.97  & 0.04          & VLBA   & no           & USNO \\
8.4   & 1999.03     & 1.57    & 1.57      & 0.31$\times$0.26 & 6.5           & VLBA   & no           & V00 \\
15    & 1995.57     &         & 0.67$^g$  & 1.2$\times$0.6   & 0.1           & VLBA   & no           & KV98 \\
15    & 1995.96     &         & 0.31$^g$  & 1.2$\times$0.6   & 0.05          & VLBA   & no           & KV98 \\
15    & 1996.31     &         & 0.36$^g$  & 1.07$\times$0.57 & 0.06          & VLBA   & no?          & USNO \\
15    & 1997.67     &         & 1.41$^g$  & 1.2$\times$0.6   & 0.2           & VLBA   & no           & KV98 \\
22    & 1982.9      & 1.38    & 1.30      & 0.1$\times$0.1   & 6             & global & $7^{\circ}$  & JB84 \\
22    & 1993.9 $-$  &         &           &                  &               & VLBA   & no           & XW98 \\
      & 1996.7 (3)  &         &           &                  &               &        &              & \\
22    & 1999.03     & 1.04    & 1.04      & 0.20$\times$0.08 & 3.2           & VLBA   & no           & V00 \\
43    & 1993.9 $-$  &         &           &                  &               & VLBA   & no           & XW98 \\
      & 1996.7 (3)  &         &           &                  &               &        &              & \\
\noalign{\smallskip}
\hline
\end{tabular*}

\begin{list}{}{}
\item[$^{\rm a}$] flux density is the sum of the fitted model component flux densities cited in the VLBI reference (where available) \
\item[$^{\rm b}$] flux density of the core \
\item[$^{\rm c}$] Gaussian model size of the core (FWHM) \
\item[$^{\rm d}$] lower limit for the core brightness temperature estimated from fitted model flux densities and angular sizes \
\item[$^{\rm e}$] EVN=European VLBI Network; SVLBI=space VLBI; VLBA=Very Long Baseline Array \
\item[$^{\rm f}$] AG95=Altschuler et al. 1995; BE81=B{\aa}{\aa}th et al. 1981; CB96=Chu et al. 1996; CU83=Cohen et al. 1983; CZ99=Chen, Zhang \& Sjouwerman 1999; FF00=Fomalont et al. 2000; GC89=Gabuzda et al. 1989; GC92=Gabuzda et al. 1992; GC96=Gabuzda \& Cawthorne 1996; JB84=Jones et al. 1984; KV98=Kellermann et al. 1998; SW97=Shen et al. 1997; USNO=US Naval Observatory ({\it http://maia.usno.navy.mil/rorf/rrfid.html}); V00=T.~Venturi et al. 2000, in preparation; XW98=Xu, Wehrle \& Marscher 1998  
\item[$^{\rm g}$] peak brightness and restoring beam size in the VLBI image were used to substitute $S$ and size, to obtain a rough estimate of $T_{\mathrm B}$, assuming that the core is unresolved \
\end{list}
\end{table*}

Flux density variations across a factor--of--50 range in radio frequencies during a 5-year 
monitoring interval were found to be correlated, suggesting that they are intrinsic to the object. 
The radio variability can qualitatively be understood in terms of adiabatically evolving structures 
in a relativistic jet (O'Dell et al. 1988).
However, Romero et al. (1997) favor refractive scintillation as the simplest explanation of 
intraday radio variability observed in AO 0235+164. Most recently,
Kraus et al. (1999) analysed contemporaneous radio and optical monitoring data taken in October 1992 
over a period of about a month. They discuss several models to explain the observations, including 
relativistic shock--fronts, precessing beams, free--free absorption by the foreground medium, 
interstellar scattering and gravitational microlensing. Since apparently none of the models 
investigated accounts for the observed variations alone, Kraus et al. suggest that the variability 
is caused by a superposition of intrinsic and propagation effects. They note that all explanations 
imply a very small intrinsic source size and require a Doppler factor substantially larger than
usual (up to 100) to be consistent with the inverse Compton brightness temperature limit of 
$\sim10^{12}$~K (Kellermann \& Pauliny-Toth, 1969).

A large number of VLBI images of AO 0235+164 have been produced over a wide range of observing 
wavelengths (from millimeters to meters). We have collected a long, albeit certainly incomplete, 
list of references in Table~1. A large fraction of the VLBI images do not show any extended 
structure apart from the very compact core, regardless of the different angular resolution 
and the actual total flux density at the observing time (e.g. Gabuzda et al. 1989, 1992, 1996; 
Xu, Wehrle \& Marscher 1998). On the other hand, several authors report faint extensions at a 
variety of position angles (PA), but mainly between the north and east (e.g. Jones et al. 1984; 
Chu et al. 1996; Shen et al. 1997). Although the components found by Chu et al. (1996) appear 
to move too fast to be identifiable between the subsequent epochs, these authors, based on the 
assumption that the components they observe are the results of outbursts at earlier epochs, 
come to an estimate of the Doppler boosting of the order $10^5$.

Several non-imaging VLBI surveys, e.g. Preston et al. (1985) at 2.3~GHz, 
Morabito et al. (1986) at 2.3 and 8.4~GHz and Moellenbrock et al. (1996) at 22~GHz, 
also found AO 0235+164 to be extremely compact. The source 
is often used as a fringe-finder and calibrator for VLBI experiments.

\section
{VSOP space VLBI observations, calibration and data reduction}

AO 0235+164 was observed in the VLBI Space Observatory Programme (VSOP, Hirabayashi et al. 1998)
at 1.6~GHz on 31 January 1999 with the HALCA satellite via the tracking stations at Tidbinbilla 
(Australia), Goldstone (USA), Usuda (Japan) and Robledo (Spain). The participating ground telescopes 
were the Australia Telescope Compact Array (ATCA), Sheshan (China), Hartebeesthoek (South Africa) 
and Noto (Italy). The 5-GHz VSOP observation took place on 1 February 1999. HALCA was tracked from 
Tidbinbilla. The three co-observing ground radio telescopes were ATCA, Sheshan and Usuda. The data 
in both experiments were recorded in left circular polarization in S2 format (Wietfeldt et al. 1996) 
with 32~MHz bandwidth. Correlation was performed at the Dominion Radio Astrophysical Observatory 
in Penticton (Canada) (Carlson et al. 1999). 

Data calibration and fringe-fitting with 5-min solution interval were done with the NRAO AIPS 
package (version 15OCT98; e.g. Cotton 1995, Diamond 1995). Time averaging, editing, 
self-calibration and imaging were done using the DIFMAP program 
(version 2.3c; Shepherd, Pearson \& Taylor 1994). 

Strong fringes were found in the 5-GHz experiment for all available antennas and time ranges. 
Measured system temperatures were available at Sheshan and Usuda for initial amplitude 
calibration. Nominal values were used for HALCA and ATCA. The calibrator source 0420$-$014 
was used to adjust the antenna gains of the ground telescopes. The total flux density of 
0420$-$014 is being monitored at the University of Michigan Radio Astronomy Observatory 
(UMRAO, {\it http:$\!$/$\!$/www.astro.lsa.umich.edu$\!$/obs$\!$/radiotel$\!$/umrao.html}). 
The source was observed in the 5-GHz VSOP-VLBA Pre-launch Survey (VLBApls, June 1996, 
Fomalont et al. 2000). Based on the similar 5-GHz total flux densities at the epochs of 
the VLBApls and our VSOP observation, and assuming that the source structure found in the 
VLBApls has not changed considerably since 1996, the ground telescope gains had to be 
corrected by factors of 1.3, 1.5 and 3.3 for ATCA, Sheshan and Usuda, respectively. 
The unusually large correction factor for Usuda is explained by the fact that the 
quantizer threshold was set incorrectly for this telescope, as noted at the correlator. 
Initial HALCA amplitude calibration could not be checked but it is known to be stable 
(Moellenbrock et al. 2000). 

The ($u,v$) coverage for the 5-GHz experiment is shown in Fig.~1. Fig.~2 shows the 
visibility amplitudes and phases as a function of projected baseline length up to 
420 million wavelengths (M$\lambda$), along with the curves representing the final 
clean-component model. The uniformly weighted image is shown in Fig.~3.


\begin{figure}
\centering
\includegraphics[clip=,bb=30pt 115pt 590pt 680pt,scale=0.34,angle=-90]
{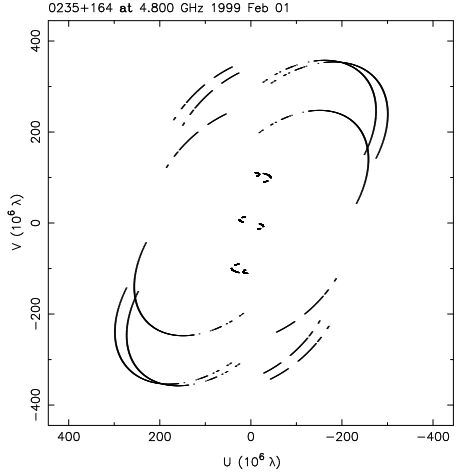}
\caption{The $(u,v)$ coverage of the 5-GHz space VLBI experiment. Participating ground antennas are ATCA, Sheshan and Usuda. 
Projected baselines to HALCA extend up to 420 M$\lambda$.}
\label{fig1} 
\end{figure}


\begin{figure}
\centering
\includegraphics[clip=,bb=80pt 20pt 580pt 740pt,scale=0.34,angle=-90]
{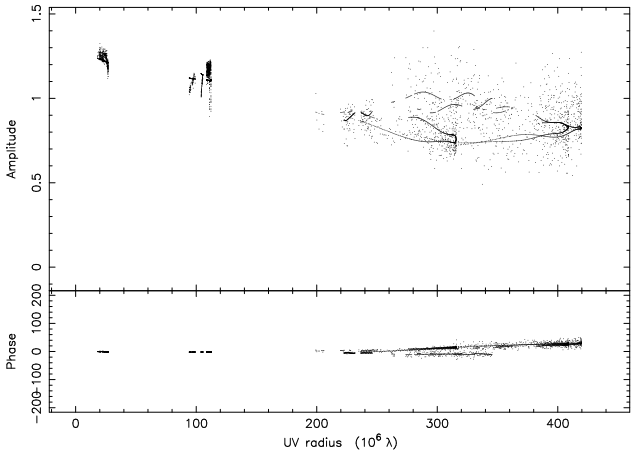}
\caption{Self-calibrated visibility amplitude (Jy) and phase (degrees) vs. projected baseline length for AO 0235+164 at 5~GHz.
Solid lines represent the clean component model used for restoring the image in Fig.~3.}
\label{fig2} 
\end{figure}


\begin{figure}
\centering
\includegraphics[clip=,bb=30pt 135pt 590pt 680pt,scale=0.34]
{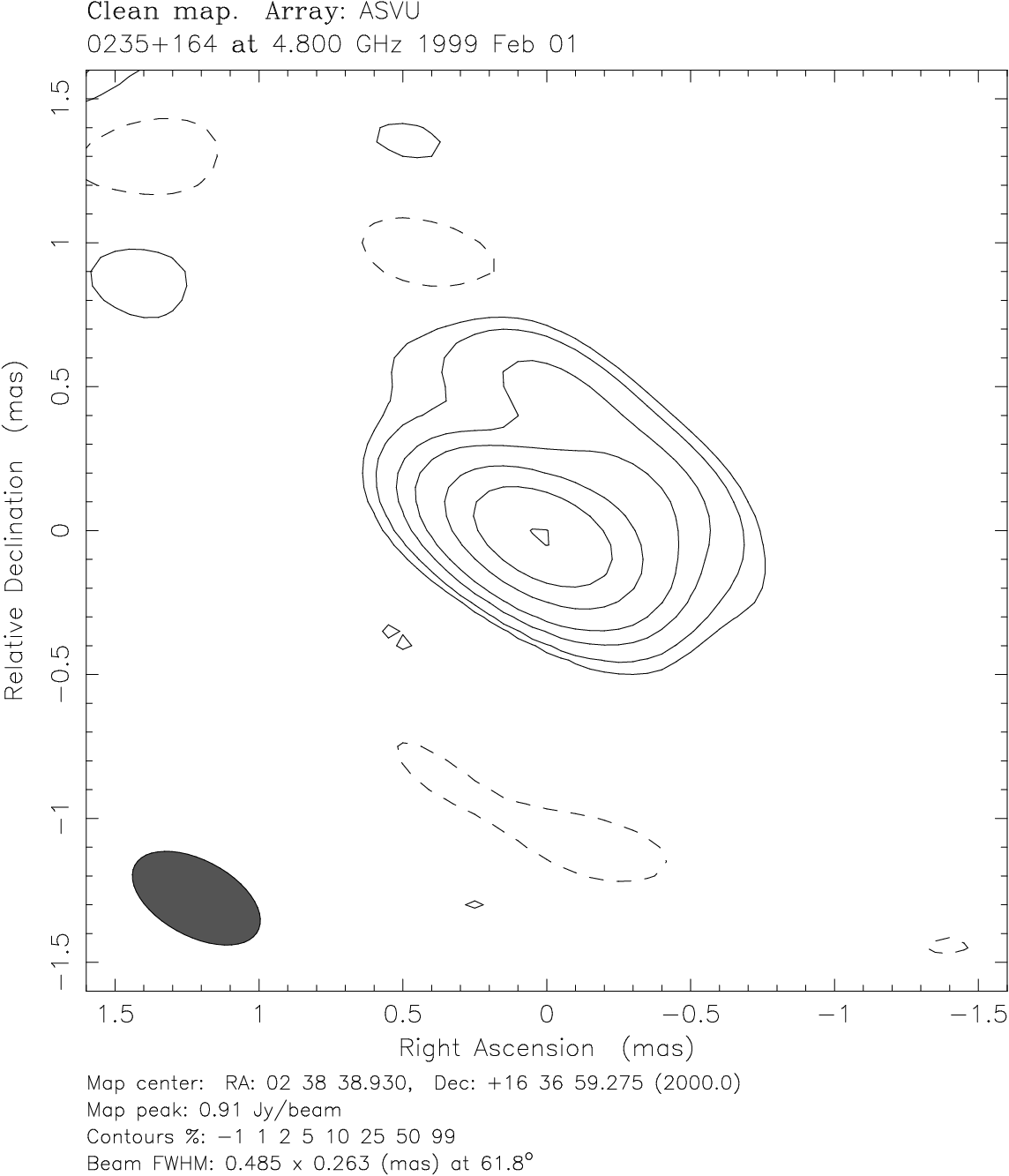}
\caption{5-GHz space VLBI image of AO 0235+164. 
Restoring beam is 0.49$\times$0.26~milliarcsecond at
PA=62$^{\circ}$, contour levels are at --1, 1, 2, 5, 10, 25, 50, 99\% of the peak brightness of 910~mJy/beam.}
\label{fig3} 
\end{figure}

In the case of the 1.6-GHz experiment, the initial amplitude calibration was based on 
the measured system temperatures and gain curves for Hartebeesthoek, Noto and Sheshan. 
Nominal system temperatures were used for ATCA and HALCA.
Fringe-fitting resulted in strong fringes for all available baselines except those to 
Sheshan, where no fringes were found. This, and the fact that the source was observed 
during a very limited common time range at the remaining antennas, made  the traditional 
hybrid mapping of AO 0235+164 impossible at this observing frequency. Model fitting to 
the visibility data in DIFMAP resulted in an elliptical Gaussian component with flux 
density of 1.44 Jy, angular size of 0.85~mas~$\times$~0.60~mas (FWHM) and major axis 
position angle of $-14^{\circ}$. Fig.~4 shows the visibility amplitudes and phases vs. 
projected baseline length. The data were edited and averaged over 30~s intervals in 
DIFMAP. The ground telescope antenna gains were again adjusted using the total flux 
density data taken simultaneously and assuming that the source is unresolved on the 
shortest baselines. The gain correction factors were 1.4, 1.4 and 1.8 for Noto, 
Hartebeesthoek and ATCA, respectively. 


\begin{figure}
\centering
\includegraphics[clip=,bb=80pt 20pt 580pt 740pt,scale=0.34,angle=-90]
{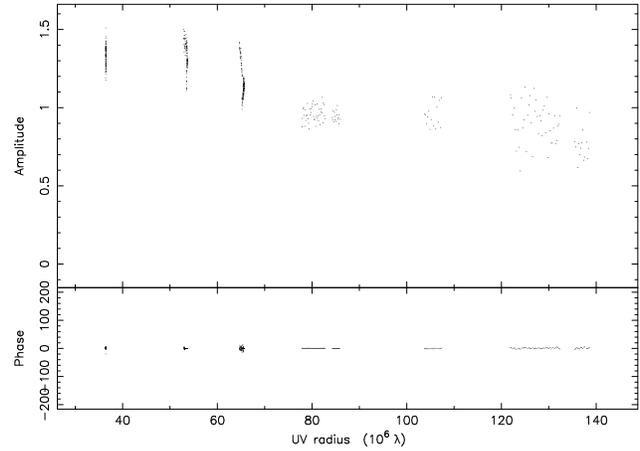}
\caption{Calibrated visibility amplitude (Jy) and phase (degrees) vs. projected baseline length for AO 0235+164 at 1.6~GHz. 
Reliable amplitude self-calibration could not be made due to the lack of four useful antennas in the array for most 
of the observing time.}  
\label{fig4} 
\end{figure}

\section
{Discussion}

\subsection{Sub-milliarcsecond structure}

At 5~GHz, the structure of AO 0235+164 on sub-milliarcsecond scale is dominated by a 
compact component, with a weak extension to approximately the north-west  (Fig.~3). 
Its position angle is not quite the same as that found in many VLBI images published 
earlier (see Table~1), including the 22-GHz ground VLBI image of Jones et al. (1984) 
which has almost exactly the same angular resolution. 
Note that earlier ground-only VLBI imaging observations led to controversial results 
on the existence and position angle of the extended features, which most likely 
indicates rapid structural variations.
However, the existence of such an extension must be treated with some caution because 
of the relatively sparse ($u,v$) coverage in our space VLBI observation. The visibility 
data of AO 0235+164 are consistent with a central point-source and an elliptical Gaussian 
component, according to model fitting in DIFMAP. The model parameters are listed in Table~2. 
Indeed, the source is remarkably compact: the correlated flux density ratio between the 
longest and shortest projected baselines is 3:5 (Fig.~2). 


\begin{table*}[t]
\small
\begin{center}
Table~2.\hspace{4pt}Fitted model parameters of the source components at 5~GHz.
Angular size upper limits and uncertainties are calculated from the image parameters, following Fomalont et al. (2000).\\
\end{center}
\vspace{6pt}
\begin{tabular*}{\textwidth}{@{\hspace{\tabcolsep}
\extracolsep{\fill}}p{7pc}cccccc} 
\hline\hline\\[-6pt]
Component 	& $S$    & $r$    & $\theta$  & $a$\,\,\,\,  & $b$ & $\Phi$      \\
          	& (mJy)  & (mas)  &($^{\circ}$)& (mas)       & (mas)      &($^{\circ}$) \\
[4pt]\hline \\[-6pt]

Core	        & 915    & 0.04   & 127       & $<0.05$      & $<0.03$ & $-$ \\ 
A	        & 319    & 0.21   & $-$57     & $0.60\pm0.02$    & $0.20\pm0.02$  & 37 \\ 
\hline
\end{tabular*}

\begin{list}{}{}
\item[]$S$: flux density, $r$: separation from the field center, $\theta$: position angle, $a$: major axis,
$b$: minor axis, $\Phi$: position angle of the major axis; position angles are measured from north through east.\\

\end{list}

\end{table*}

\subsection{Source spectrum}

At the time of our observations, the total flux density of AO 0235+164 was declining 
according to the data from the UMRAO long--term monitoring program at three different 
radio frequencies (Aller et al. 1999). The total flux density data at 4.8~GHz are shown 
in Fig.~5, indicating the epoch of our observations. The quasi-single epoch source spectrum 
(Fig.~6) is constructed from measurements at UMRAO and the Arecibo Observatory on 30 January 1999. 
The flux density over a range spanning an order of magnitude in frequency (from 1.175 to 14.5~GHz) 
is roughly constant. It can be well described with a power-law spectral index, 
$\alpha=-0.02\pm0.03$ ($S\sim\nu^{\alpha}$). Although there are no measurements available at 
lower frequencies, a spectral turnover at $\sim2$~GHz may be suspected.

It should be noted that the relatively large ground radio telescope gain correction factors 
used for amplitude calibration of our space VLBI data lead to correlated flux densities of 
$\sim1.5$~Jy on the shortest baselines. This is in agreement with the contemporaneously 
measured total flux densities and the well-known compact nature of AO 0235+164 which makes 
this source a widely used VLBI fringe-finder and calibrator.


\begin{figure}
\centering
\includegraphics[clip=,bb=60pt 20pt 580pt 740pt,scale=0.34,angle=-90]
{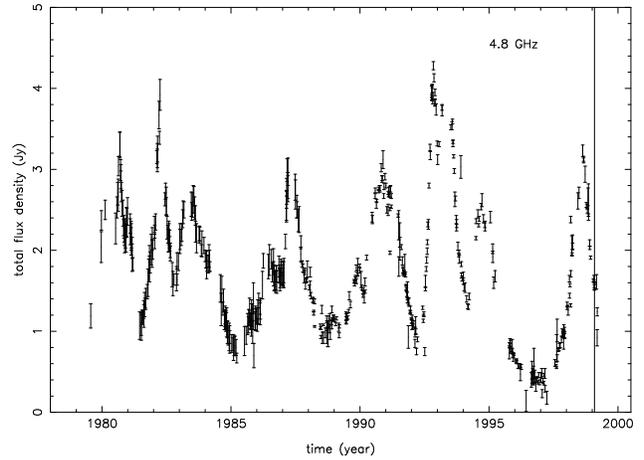}
\caption{Total flux density of AO 0235+164 at 4.8~GHz measured at the University of Michigan Radio Observatory (UMRAO) 
26-m antenna since 1979. The vertical line indicates the epoch of our observations.}  
\label{fig5} 
\end{figure}


\begin{figure}
\centering
\includegraphics[clip=,bb=80pt 20pt 580pt 740pt,scale=0.34,angle=-90]
{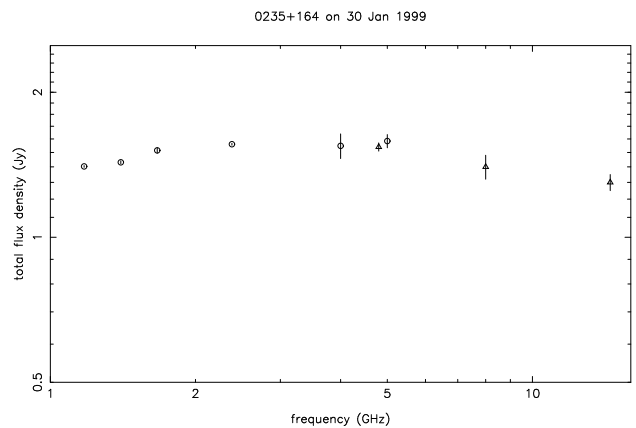}
\caption{Total flux density of AO 0235+164 at nine different frequencies measured nearly contemporaneously at the 
Arecibo Observatory (open circles) and at the UMRAO (triangles). Values at lower frequencies are taken on 
30 January 1999. Flux densities at 8 and 14.5~GHz are linearly interpolated using the closest in time UMRAO data.}\label{fig6} 
\end{figure}

\subsection{Brightness temperature} 

A lower limit for the rest--frame brightness temperature of the source can be estimated as
\begin{equation}
T_{\mathrm B} [K] =  1.22 \times 10^{12}  \frac{S}{\theta_{maj}\theta_{min}} \frac{1+z}{\nu^2},
\end{equation}
where $S$ is the component flux density in Jy, $\theta_{maj}$ and $\theta_{min}$ are the major 
and minor axes, respectively, in mas (assuming optically thick Gaussian brightness distribution), 
$\nu$ is the observing frequency in GHz and $z$ is the redshift. For the core of AO 0235+164, 
at the redshift of $z=0.94$, we obtain $T_{\mathrm B}>5.8\times10^{13}$~K using the source model 
from the 5-GHz experiment (Table~2). This is a direct evidence for a brightness temperature 
exceeding the inverse Compton limit and can be interpreted, in particular, as a result of 
relativistic motion in the source.  

Earlier brightness temperature estimates 
for AO 0235+164 were based on e.g. 22-GHz VLBI survey 
data ($T_{\mathrm B}>2\times10^{12}$~K, Moellenbrock et al. 1996), 5-GHz ground-based VLBI 
imaging ($T_{\mathrm B}=5.3\times10^{12}$~K, Shen et al. 1997) and variability 
($T_{\mathrm B}=7\times10^{17}$~K for $\lambda=20$~cm, Kraus et al. 1999). 
We have made an attempt to compile a list of brightness temperature lower limit estimates 
using the VLBI imaging experiments cited in Table~1, according to Eq.~1. Calculated $T_{\mathrm B}$ 
values are based on fitted Gaussian core model component flux densities and 
angular sizes given in the references where available. The brightness temperature was estimated 
using image parameters (peak brightness and restoring beam size, assuming that the source is not 
resolved) where only images were available. The values as a function of time are plotted in Fig.~7 
using the same time scale as for the total flux density monitoring data (Fig.~5). The estimates of 
$T_{\mathrm B}$ are lower limits, the time sampling is quite sparse, and the values were calculated 
from VLBI experiments made at different frequencies, and thus with different sensitivities, to 
determine the brightness temperature. Nevertheless, there is an indication of a correlation between 
the long term behaviour of $T_{\mathrm B}$ and measured total flux densities (Fig.~8). The correlation 
cannot arise from the dependence of $T_{\mathrm B}$ on $S$ alone, since source size is also highly 
variable (see Table 1) and $T_{\mathrm B}$ varies in a range spanning four orders of magnitude (Fig. 7). 
This possible correlation is qualitatively consistent with the picture that each flux density outburst 
marks the birth of a new, relativistic component which later fades away and the angle between the 
direction of motion and the line of sight increases (Chu et al. 1996). As the jet direction in 
BL Lac objects is thought to be oriented very close to the line of sight, this change causes 
significant decrease in the Doppler boosting factor.


\begin{figure}
\centering
\includegraphics[clip=,bb=60pt 20pt 580pt 740pt,scale=0.34,angle=-90]
{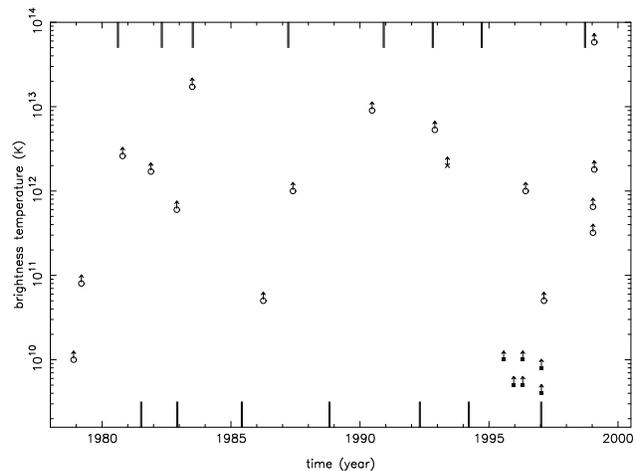}
\caption{Estimated lower limits of brightness temperature of AO 0235+164 vs. time. Open circles and filled boxes 
indicate estimates based on fitted VLBI model components and image parameters, respectively 
(see footnotes to Table~1 for explanation). An estimate of Moellenbrock et al. (1996) using non--imaging 
VLBI survey data is shown with a cross. Times of local maxima and minima in the 4.8-GHz total flux density 
curve (Fig.~5) are indicated with large tick marks at the top and bottom of the plot, respectively.} 
\label{fig7}
\end{figure}


\begin{figure}
\centering
\includegraphics[clip=,bb=80pt 20pt 580pt 740pt,scale=0.34,angle=-90]
{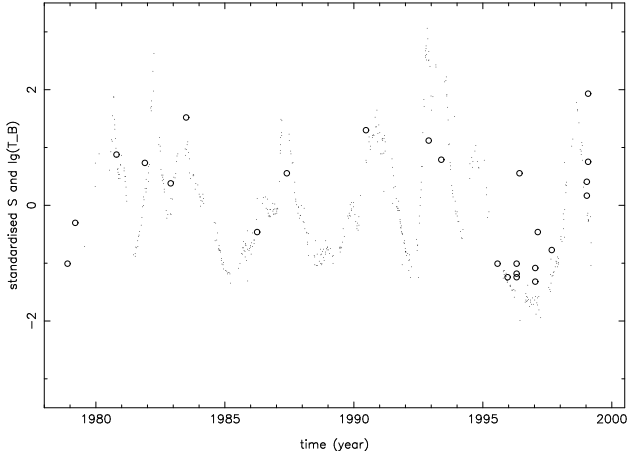}
\caption{Standardised values of the 4.8-GHz total flux density ($S_0$, dots; Fig.~5) and of the logarithms of 
brightness temperature lower limits ($T_{\mathrm B}$, circles; Fig.~7) are plotted together as a function 
of time, to illustrate the apparent correlation between the two data sets. The correlation coefficient 
calculated using the closest-in-time $S_0$ and $T_{\mathrm B}$ values is 0.7.}
\label{fig8} 
\end{figure}

\section{Conclusions}

Our 5-GHz VSOP space VLBI experiment shows the BL~Lac object AO 0235+164 to be very compact on 
sub-milliarcsecond angular scales. The rest-frame brightness temperatures estimated from the 
source angular size and observed flux density clearly exceed $10^{12}$~K which can be explained by 
strong Doppler boosting. The brightness temperature $T_{\mathrm B}>5.8\times10^{13}$~K obtained 
from the 5-GHz data is the highest value found with VSOP observations to date (cf. Lovell et al. 2000). 
This implies a Doppler factor of $\sim100$, in good agreement with the results of several recent 
studies (e.g. Kraus et al. 1999, Fujisawa et al. 1999), although the source total flux density was 
relatively low at the time of our observations.

There is an indication of an extended component to N-NW, at the position angle about $90^{\circ}$ 
away from what is sometimes seen in ground VLBI images. Due to the possibly very small jet angle 
with respect to the line of sight (e.g. $\phi\leq0^{\circ}\hspace{-4.5pt}.\hspace{0.5pt}35$ is 
estimated by Fujisawa et al. 1999), small deviations in the intrinsic direction of the jet ejection 
could cause large changes in the observed jet position angle. 

Further high resolution VLBI monitoring observations, either with space VLBI and/or ground-based 
VLBI at higher frequencies ($\nu\geq15$~GHz), would be needed to identify possible components ejected 
after major outbursts occuring every 3--5 years.

\par
\vspace{1pc}\par
We gratefully acknowledge the VSOP Project, which is led by the Institute of Space and Astronautical 
Science (Japan) in cooperation with many organizations and radio telescopes around the world. 
The Arecibo Observatory is operated by Cornell University under cooperative agreement with the National 
Science Foundation. This research has made use of data from the University of Michigan Radio Astronomy 
Observatory which is supported by funds from the University of Michigan and the National Science Foundation. 
This research has made use of the United States Naval Observatory (USNO) Radio Reference Frame Image Database. 
SF acknowledges financial support received from the Hungarian Space Office, the Netherlands Organization for 
Scientific Research and the Hungarian Scientific Research Fund (grant no. N31721 \& T031723). 
We thank Tiziana Venturi for providing us with their 8.4 and 22-GHz VLBA imaging results prior to publication.

\section*{References}
\small

\re
Abraham G.R., Crawford C.S., Merrifield M.R., Hutchings J.B., McHardy I.M.\ 1993, ApJ 415, 101

\re 
Aller M.F., Aller H.D., Hughes P.A., Latimer G.A.\ 1999, ApJ 512, 601

\re
Altschuler D.R., Gurvits L.I., Alef W., Dennison B., Graham D., Trotter A.S., Carson J.E.\ 1995, A\&AS 114, 197

\re
B{\aa}{\aa}th L.B., Elgered G.K., Lundqvist G., Graham D., Weiler K.W., Seielstad G.A., Tallqvist S., 
Schilizzi R.T.\ 1981, A\&A 96, 316

\re
Burbidge E.M., Beaver E.A., Cohen R.D., Junkkarinen V.T., Lyons R.W.\ 1996, AJ 112, 2533

\re
Carlson B.R., Dewdney P.E., Burgess T.A., Casorso R.V., Petrachenko W.T., Cannon W.H.\ 1999, PASP 111, 1025  

\re
Chen Y.J., Zhang F.J., Sjouwerman L.O.\ 1999, New\ Astron.\ Rev. 43, 707 

\re
Chu H.S., B{\aa}{\aa}th L.B., Rantakyr\"o F.T., Zhang F.J., Nicolson G.\ 1996, A\&A 307, 15

\re
Cohen M.H., Unwin S.C., Lind K.R., et al.\ 1983, ApJ 272, 383

\re
Cohen R.D., Smith H.E., Junkkarinen V.T., Burbidge E.M.\ 1987, ApJ 318, 577

\re 
Cotton W.D.\ 1995, in: Very Long Baseline Interferometry and the VLBA, eds. Zensus J.A., Diamond P.J., 
Napier P.J., ASP Conference Series 82 (ASP, San Francisco), 189

\re 
Diamond P.J.\ 1995, in: Very Long Baseline Interferometry and the VLBA, eds. Zensus J.A., Diamond P.J., 
Napier P.J., ASP Conference Series 82 (ASP, San Francisco), 227

\re
Fomalont E.B., Frey S., Paragi Z., Gurvits L.I., Scott W.K., Taylor A.R.,Edwards P.G., Hirabayashi H.\ 2000, 
ApJS, in press

\re
Fujisawa K., Kobayashi H., Wajima K., Hirabayashi H., Kameno S., Inoue M.\ 1999, PASJ 51, 537

\re
Gabuzda D.C., Cawthorne T.V.\ 1996, MNRAS 283, 759

\re
Gabuzda D.C., Cawthorne T.V., Roberts D.H., Wardle J.F.C.\ 1989, ApJ 347, 701

\re
Gabuzda D.C., Cawthorne T.V., Roberts D.H., Wardle J.F.C.\ 1992, ApJ 388, 40

\re
Hirabayashi H., Hirosawa H., Kobayashi H., et al.\ 1998, Science 281, 1825

\re
Jones D.L., B{\aa}{\aa}th L.B., Davis M.M., Unwin S.C.\ 1984, ApJ 284, 60

\re
Kellermann K.I., Pauliny-Toth I.I.K.\ 1969, ApJ 155, L71

\re
Kellermann K.I., Vermeulen R.C., Zensus J.A., Cohen M.H.\ 1998, AJ 115, 1295

\re
Kraus A., Quirrenbach A., Lobanov A.P., et al.\ 1999, A\&A 344, 807

\re
Lovell J.E.J., Horiuchi S., Moellenbrock G., et al.\ 2000, in: Astrophysical Phenomena Revealed by Space VLBI, 
eds. Hirabayashi H., Edwards P.G., Murphy D.W. (ISAS, Sagamihara), 183

\re
Madejski G., Takahashi T., Tashiro M., Kubo H., Hartman R., Kallman T., Sikora M.\ 1996, ApJ 459, 156

\re
Moellenbrock G.A., Fujisawa K., Preston R.A., et al.\ 1996, AJ 111, 2174

\re
Moellenbrock G.A., Lovell J., Horiuchi S., et al.\ 2000, in: Astrophysical Phenomena Revealed by Space VLBI, 
eds. Hirabayashi H., Edwards P.G., Murphy D.W. (ISAS, Sagamihara), 177

\re
Morabito D.D., Neill A.E., Preston R.A., Linfield R.P., Wehrle A.E., Faulkner J. \ 1986, AJ 91, 1038

\re
Nilsson K., Charles P.A., Pursimo T., Takalo L.O., Sillanp\"a\"a A., Teerikorpi P.\ 1996, A\&A 314, 754

\re
O'Dell S.L., Dennison B., Broderick J.J., et al.\ 1988, ApJ 326, 668

\re
Preston R.A., Morabito D.D., Williams J.G., Faulkner J., Jauncey D.L., Nicolson G.D.\ 1985, AJ 90, 1599

\re
Rabbette M., McBreen B., Steel S., Smith N.\ 1996, A\&A 310, 1

\re
Romero G.E., Combi J.A., Benaglia P., Azc\'arate I.N., Cersosimo J.C., Wilkes L.M.\ 1997, A\&A 326, 77

\re
Saust A.B.\ 1992, A\&A 266, 101

\re
Shen Z.Q., Wan T.S., Moran J.M., et al.\ 1997, AJ 114, 1999

\re 
Shepherd M.C., Pearson T.J., Taylor G.B.\ 1994, BAAS 26, 987

\re
Spinrad H., Smith H.E.\ 1975, ApJ 201, 275

\re
Stickel M., Fried J.W., K\"uhr H.\ 1988, A\&A 198, L13

\re
Takalo L.O., Kidger M.R., De Diego J.A., Sillanp\"a\"a A., Nilsson K.\ 1992, AJ 104, 40

\re
von Montigny C., Bertsch D.L., Chiang J. et al.\ 1995, ApJ 440, 525 

\re
Wietfeldt R.D., Baer D., Cannon W.H., Feil G., Jakovina R., Leone P., Newby P., Tan H.\ 1996, IEEE Trans. Instrum. Meas. 45, 923

\re
Xu W., Wehrle A.E., Marscher A.P.\ 1998, in: Proc. IAU Colloquium 164, Radio Emission from Galactic and Extragalactic 
Compact Sources, eds. Zensus J.A., Taylor G.B., Wrobel J.M., ASP Conference Series 144 (ASP, San Francisco), 175

\label{last}

\end{document}